\documentclass[useAMS,usenatbib,usegraphicx]{mn2e}

\usepackage{color}
\usepackage{natbib, aas_macros}
\title[What if LIGO's GWs are strongly lensed?]{What if LIGO's
   gravitational wave detections are strongly lensed
  by massive galaxy clusters?}

\author[G.\ P.\ Smith et al.]
       {Graham P.\ Smith,$\!^1$\thanks{E-mail: gps@star.sr.bham.ac.uk}
        Mathilde Jauzac,$\!^{2,3,4,5}$
        John Veitch,$\!^{1,6,7}$
        Will M.\ Farr,$\!^{1,7}$\newauthor
        Richard Massey,$\!^2$
        Johan Richard$^8$\\\\
$^1$ School of Physics and Astronomy, University of
Birmingham, Birmingham, B15 2TT, England\\
$^2$ Centre for Extragalactic Astronomy, Department of Physics, 
Durham University, Durham DH1 3LE, England\\
$^3$ Institute for Computational Cosmology, Durham University, 
South Road, Durham DH1 3LE, England\\
$^4$ Astrophysics and Cosmology Research Unit, 
School of Mathematical Sciences, University of KwaZulu-Natal, 
Durban 4041, South Africa\\
$^5$ Laboratoire d'Astrophysique \'Ecole Polytechnique
F\'ed\'erale de Lausanne (EPFL) Observatoire de Sauverny
CH-1290 Versoix, Switzerland\\
$^6$ School of Physics and Astronomy, University of Glasgow, G12 8QQ, Scotland\\
$^7$ Birmingham Institute of Gravitational Wave Astronomy, University of
Birmingham, Birmingham, B15 2TT, England\\
$^8$ CRAL, Observatoire de Lyon, Universit\'e Lyon 1, 9 Avenue
Ch.\ Andr\'e, 69561 Saint Genis Laval Cedex, France\\
       }

\begin{document}

\date{\today}

\pagerange{\pageref{firstpage}--\pageref{lastpage}} \pubyear{2017}

\maketitle

\label{firstpage}

\newcommand{\simgt}{\lower.5ex\hbox{$\; \buildrel > \over \sim \;$}}
\newcommand{\simlt}{\lower.5ex\hbox{$\; \buildrel < \over \sim \;$}}
\newcommand{\tp}{\hspace{-1mm}+\hspace{-1mm}}
\newcommand{\tm}{\hspace{-1mm}-\hspace{-1mm}}
\newcommand{\gIII}{I\hspace{-.3mm}I\hspace{-.3mm}I}
\newcommand{\bmf}[1]{\mbox{\boldmath$#1$}}
\def\bbeta{\mbox{\boldmath $\beta$}}
\def\btheta{\mbox{\boldmath $\theta$}}
\def\bnabla{\mbox{\boldmath $\nabla$}}
\def\bk{\mbox{\boldmath $k$}}
\newcommand{\cA}{{\cal A}}
\newcommand{\cD}{{\cal D}}
\newcommand{\cF}{{\cal F}}
\newcommand{\cG}{{\cal G}}
\newcommand{\trQ}{{\rm tr}Q}
\newcommand{\Real}[1]{{\rm Re}\left[ #1 \right]}
\newcommand{\paren}[1]{\left( #1 \right)}
\newcommand{\red}{\textcolor{red}}
\newcommand{\blue}{\textcolor{blue}}
\newcommand{\norv}[1]{{\textcolor{blue}{#1}}}
\newcommand{\norvnew}[1]{{\textcolor{red}{#1}}}

\def\hkpc{\mathrel{h^{-1}{\rm kpc}}}
\def\hMpc{\mathrel{h^{-1}{\rm Mpc}}}
\def\Mpc{\mathrel{\rm Mpc}}
\def\Mvir{\mathrel{M_{\rm vir}}}
\def\cvir{\mathrel{c_{\rm vir}}}
\def\rvir{\mathrel{r_{\rm vir}}}
\def\Dvir{\mathrel{\Delta_{\rm vir}}}
\def\rsc{\mathrel{r_{\rm sc}}}
\def\rhoc{\mathrel{\rho_{\rm crit}}}
\def\Msol{\mathrel{M_\odot}}
\def\hMsol{\mathrel{h^{-1}M_\odot}}
\def\h70Msol{\mathrel{h_{70}^{-1}M_\odot}}
\def\ergs{\mathrel{\rm erg\,s^{-1}}}
\def\Mgas{\mathrel{M_{\rm gas}}}
\def\Mhse{\mathrel{M_{\rm HSE}}}
\def\Mp{\mathrel{M_{\rm Planck}}}
\def\Mwl{\mathrel{M_{\rm WL}}}
\def\Mfh{\mathrel{M_{500}}}
\def\rfh{\mathrel{r_{500}}}
\def\Tx{\mathrel{T_X}}
\def\Om{\mathrel{\Omega_{\rm M}}}
\def\Ol{\mathrel{\Omega_\Lambda}}
\def\keV{\mathrel{\rm keV}}
\def\kpc{\mathrel{\rm kpc}}
\def\kms{\mathrel{\rm km\,s^{-1}}}
\def\ls{\mathrel{\hbox{\rlap{\hbox{\lower4pt\hbox{$\sim$}}}\hbox{$<$}}}}
\def\gs{\mathrel{\hbox{\rlap{\hbox{\lower4pt\hbox{$\sim$}}}\hbox{$>$}}}}
\def\ds{\mathrel{D_{\rm S}}}
\def\dls{\mathrel{D_{\rm LS}}}
\def\dsi{\mathrel{D_{{\rm S},i}}}
\def\dlsi{\mathrel{D_{{\rm LS},i}}}
\def\dsj{\mathrel{D_{{\rm S},j}}}
\def\dlsj{\mathrel{D_{{\rm LS},j}}}
\def\dsi{\mathrel{D_{{\rm S},i}}}
\def\zs{\mathrel{z_{S}}}
\def\ks{\mathrel{\rm ksec}}
\def\betaP{\mathrel{\beta_{\rm P}}}
\def\betaX{\mathrel{\beta_{\rm X}}}
\newcommand{\perGpcyr}{\ensuremath{\mathrm{Gpc}^{-3} \, \mathrm{yr}^{-1}}}
\newcommand{\be}{\begin{equation}}
\newcommand{\ee}{\end{equation}}
\newcommand{\ba}{\begin{eqnarray}}
\newcommand{\ea}{\end{eqnarray}}

\addtolength\topmargin{-12mm}

\begin{abstract}
  Motivated by the preponderance of so-called ``heavy black holes'' in
  the binary black hole (BBH) gravitational wave (GW) detections to
  date, and the role that gravitational lensing continues to play in
  discovering new galaxy populations, we explore the possibility that
  the GWs are strongly-lensed by massive galaxy clusters.  For
  example, if one of the GW sources were actually located at $z=1$,
  then the rest-frame mass of the associated BHs would be reduced by a
  factor $\sim2$.  Based on the known populations of BBH GW sources
  and strong-lensing clusters, we estimate a conservative lower limit
  on the number of BBH mergers detected per detector year at
  LIGO/Virgo's current sensitivity that are multiply-imaged, of
  $R_{\rm detect}\simeq10^{-5}\,{\rm yr}^{-1}$.  This is equivalent to
  rejecting the hypothesis that one of the BBH GWs detected to date
  was multiply-imaged at $\ls4\sigma$.  It is therefore unlikely, but
  not impossible that one of the GWs is multiply-imaged.  We identify
  three spectroscopically confirmed strong-lensing clusters with well
  constrained mass models within the 90\% credible sky localisations
  of the BBH GWs from LIGO's first observing run.  In the event that
  one of these clusters multiply-imaged one of the BBH GWs, we predict
  that $20-60\%$ of the putative next appearances of the GWs would be
  detectable by LIGO, and that they would arrive at Earth within three
  years of first detection.
\end{abstract}


\makeatletter

\def\doi{\begingroup
  % The following isn't just \dospecials, because that includes \ , \{, and \}
  \let\do\@makeother \do\\\do\$\do\&\do\#\do\^\do\_\do\%\do\~
  \@ifnextchar[%]
    {\@doi}
    {\@doi[]}}
\def\@doi[#1]#2{%
  \def\@tempa{#1}%
  \ifx\@tempa\@empty
    \href{http://dx.doi.org/#2}{doi:#2}%
  \else
    \href{http://dx.doi.org/#2}{#1}%
  \fi
  \endgroup
}

%
\def\eprint#1#2{%
  \@eprint#1:#2::\@nil}
\def\@eprint@arXiv#1{\href{http://arxiv.org/abs/#1}{{\tt arXiv:#1}}}
\def\@eprint@dblp#1{\href{http://dblp.uni-trier.de/rec/bibtex/#1.xml}{dblp:#1}}
\def\@eprint#1:#2:#3:#4\@nil{%
  \def\@tempa{#1}%
  \def\@tempb{#2}%
  \def\@tempc{#3}%
  \ifx\@tempc\@empty
    \let\@tempc\@tempb
    \let\@tempb\@tempa
  \fi
  \ifx\@tempb\@empty
    \def\@tempb{arXiv}%
  \fi
  \@ifundefined{@eprint@\@tempb}
    {\@tempb:\@tempc}
    {
      \expandafter\expandafter\csname @eprint@\@tempb\endcsname\expandafter{\@tempc}}%
}

%
\def\mniiiauthor#1#2#3{%
  \@ifundefined{mniiiauth@#1}
    {\global\expandafter\let\csname mniiiauth@#1\endcsname\null #2}
    {#3}}

\makeatother


\section{Introduction}\label{sec:intro}

\noindent
Strong gravitational lensing -- i.e.\ multiple-imaging of a single
galaxy -- by massive galaxy clusters plays an invaluable role in
discovering and studying new populations of objects at high redshift
\citep[e.g.][]{Mellier91, Franx97, Ellis01, Smith02b, Kneib04b,
  Stark2007, Willis2008, Wardlow2013, Zheng2014, Bouwens2014,
  Atek2015, McLeod2015}.  Indeed, gravitational magnification by
massive clusters -- albeit not multiple-imaging -- was instrumental in
the first detections of sub-mm galaxies \citep{Smail97,Ivison98}.
More recent work has also shown that an efficient hunting ground for
strongly-lensed sub-mm galaxies is the population with the brightest
apparent fluxes \citep{Wardlow2013}.  Theoretical considerations also
underline the important role of strong-lensing by galaxy clusters in
discovering new high-redshift populations, because clusters dominate
the lensing cross-section at the large gravitational magnifications
associated with multiple-imaging ($|\mu|>10$; see Figure 5 of
\citealt{Hilbert2008}).  As the LIGO/Virgo interferometers have begun
to detect a new population of objects -- mergers of binary compact
objects \citep{BBH-O1, Abbott2017-GW170104, Abbott2017-GW170814}, it
is therefore natural to speculate on whether gravitational lensing
played a role in any of these detections.

Strong-lensing of GWs had been considered by numerous authors in
advance of the advent of direct GW detections
\citep{Wang1996,Takahashi03, Takahashi04, Seto04, Varvella04,
  Sereno2010, Sereno2011, Piorkowska2013, Biesiada2014}.  In
particular, the degeneracy between the luminosity distance to and
thus source-frame mass of a GW source, and any gravitational
magnification suffered by the source noted by \citet{Wang1996} is
interesting in light of the reported BH masses thus far \citep[see
  also][]{Dai2017}.  It is intriguing that six of the ten BHs reported
to date by LIGO/Virgo have rest-frame masses of $\gs20M_\odot$
\citep{BBH-O1, Abbott2017-GW170104, Abbott2017-GW170814}, and thus are
more massive than the most massive stellar mass BHs observed in the
local universe \citep{Farr2011}.  Whilst plausible astrophysical
interpretations of these ``heavy'' BHs do exist
\citep{Interpretation150914, Stevenson2017}, it is also possible that
the large detector-frame masses arise from lower mass sources at
higher redshift that have been gravitationally magnified.  Ignoring
this gravitational magnification would cause the redshift of the GW
sources to be underestimated, the BH masses to be overestimated, and
raise the possibility of detecting the same object again in the
future.

The GW detections have stimulated a flurry of articles on
strong-lensing of GWs, discussing the effect of lens magnification on
the detectability of GWs \citep{Dai2017}, forecast event rates
including the effects of strong-lensing by galaxies \citep{Ng2017},
relative arrival times of GW and electromagnetic (EM) signals
\citep{Takahashi2017}, prospects for measuring the speed of GWs
\citep{Collett2017, Fan2017}, and the impact of strong-lensing on
cosmography \citep{Baker2017, Liao2017}.  In this article we
investigate the probability that one or more of the GW sources
detected to date was strongly-lensed by a massive galaxy cluster.  We
take an empirical/observational approach, in that after many years of
investment by the cluster strong-lensing and LIGO/Virgo communities,
we are finally able to consider populations of observed lenses and GW
sources.  In essence, we clarify whether the perspective of an
extragalactic observer that ``surely the first detections have
benefitted from lensing'' is valid.  We also compare the sky
localisations of GW sources to the celestial coordinates of known and
spectroscopically confirmed cluster strong lenses, to identify
candidate lensing clusters that might have magnified our view of the
GWs.  For these candidate clusters we use detailed and well
constrained models of the cluster cores to answer the question: ``what
if LIGO's GW detections are strongly lensed by massive galaxy
clusters?'' -- i.e. when would we see the same sources again, and is
detection possible?

In Section~\ref{sec:glgw} we review how strong gravitational lensing
modifies GW signals and estimate the probability of strong-lensing.
We then identify the candidate cluster lenses in
Section~\ref{sec:lenses}, and describe our lensing calculations in
Section~\ref{sec:calc}.  In Section~\ref{sec:summary} we summarise and
discuss our main results.  We assume $H_0=70\kms\Mpc^{-1}$, $\Om=0.3$
and $\Ol=0.7$.


\section{Strong lensing of gravitational waves}\label{sec:glgw}

\begin{figure}
  \centerline{
    \includegraphics[width=0.75\hsize,angle=-90]{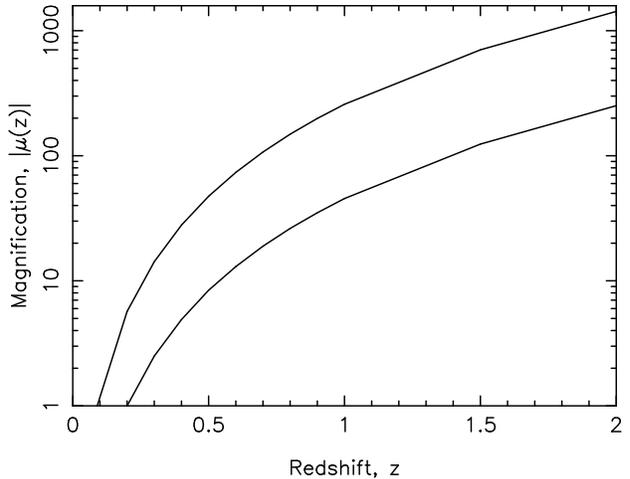}
  }
  \caption{Gravitational magnification, $\mu_{\rm GW}(z)$, required to
    modify the inferred luminosity distance to a GW source, based on
    GW150914, GW151226, GW170814 (upper curve) and LVT151012, GW170104
    (lower curve) as a function of redshift.}
  \label{fig:mugw}
\end{figure}

We consider merging compact objects as point sources of gravitational
and EM radiation.  The effect of strong gravitational lensing on short
wavelength radiation from point sources is to modify their flux
(magnification) and arrival time, yet leave their frequency unaltered
\citep[e.g.][]{Schneider92}.  In the case of GW sources, the ``flux''
manifests itself as the amplitude of the strain signal detected by the
interferometer.  In fact, the interpretation of the strain amplitude,
$A$, is degenerate to the gravitational magnification, $\mu$, and the
luminosity distance to the source, $D_{\rm L}$, as follows:
$A\propto\sqrt{|\mu|}/D_{\rm L}$ \citep{Wang1996}.  The gravitational
magnification required to reinterpret a GW source as being
strongly-lensed and thus at a higher redshift, $z$, than originally
inferred is therefore given by $\mu_{\rm GW}=[D_{\rm L}(z)/D_{\rm
    L,\mu=1}]^2$, where $D_{\rm L,\mu=1}$ is the luminosity distance
inferred assuming $\mu=1$ (Figure~\ref{fig:mugw}).

The BBH GW sources detected to date have been interpreted, assuming no
gravitational lensing, as lying at redshifts of $z\sim0.1-0.2$
\citep{BBH-O1,Abbott2017-GW170104,Abbott2017-GW170814}.  To
reinterpret a source initially identified at $z\simeq0.1$ (GW150914,
GW151226, GW170814) as actually being at $z=1$ requires $\mu_{\rm
  GW}\simeq200$, and to reinterpret sources initially identified at
$z\simeq0.2$ (LVT151012, GW170104) as actually being at $z=1$ requires
$\mu_{\rm GW}\simeq45$.  Increasing the redshift of the sources in
this way would also lead to a reduction in the inferred rest frame
masses by a factor $(1+z)$.  The masses of sources identified at
$z\simeq0.1$ would reduce by a factor $\sim1.8$ and at $z\simeq0.2$ by
a factor $\sim1.7$ if they are actually located at $z=1$.  Typical
strongly-lensed galaxies suffer gravitational magnifications of
$\mu\sim10-50$ \cite[e.g.][]{Richard10}, i.e.\ generally less than
those discussed here.  Nevertheless, very high magnifications are
physically possible because the physical region from which GWs emerge
is $\sim100{\rm km}$ in size.  It is therefore possible for a GW
source to be very closely aligned with the caustic of a gravitational
lens, and thus achieve a high magnification value \citep{Ng2017}.
This is not the case for a galaxy with a typical size of
$\sim1-10\kpc$.  In addition to revised redshifts and rest frame
masses, strongly-lensed (hereafter multiply-imaged) GW source will
arrive at Earth on multiple occasions due to the existence of several
stationary points on the Fermat surface that describes the arrival
time at Earth.  The time delay between multiple images created by
strong-lensing by galaxy clusters can be as short as a few days and as
long as $\sim10$ years \citep[e.g.][]{Jauzac2016}.

We now consider how likely it is that a GW source is multiply-imaged,
and write the number of multiply-imaged GW sources in the universe per
detector year as:
\begin{equation}
R=\int_0^{z_{\rm max}}dz_{\rm L}\int_{z_{\rm L}}^{\infty}\frac{dz}{1+z}\,\,\frac{dV_\mu}{dz\,dz_{\rm L}}\,\frac{dn}{dV\,dt},
\label{eqn:rate}
\end{equation}
where $z_{\rm L}$ is the redshift of the lens, $z$ is the actual
redshift of the GW source, $dn/dV/dt$ is the number of sources per
unit comoving volume per source-frame year, and $V_\mu$ is the
comoving volume that is magnified by $\mu=\mu_{\rm GW}$.  Note that
the sensitivity of the GW detectors is incorporated within $V_\mu$,
via the requirement for a given level of magnification to render a
distant source detectable at a given detector sensitivity.  In this
article we focus on multiple imaging of BBH mergers by known
strong-lensing clusters.  This is motivated by the absence of EM
counterparts and thus absence of precise sky localisations for BBH
mergers to date, their ``unlensed'' luminosity distances being
sufficiently large as to not require very extreme values of $\mu_{\rm
  GW}$, that clusters dominate the lensing cross-sections at high
magnification \citep{Hilbert2008}, and the availability of detailed
models of the known cluster lenses
(Sections~\ref{sec:lenses}~\&~\ref{sec:calc}).  We therefore adapt
Equation~\ref{eqn:rate} to estimate $R_{\rm detect}$, the number of
BBH mergers detected per detector year that are multiply-imaged by a
known and spectroscopically confirmed strong-lensing cluster:
\begin{equation}
R_{\rm detect}\simeq N_{\rm SL}\int_{z_{\rm L}}^{\infty}\frac{dz}{1+z}\,\,\left[\frac{dV_\mu}{dz}\right]_{\rm CL}\,\frac{dn_{\rm BBH}}{dV\,dt},\label{eqn:ratedetect}
\end{equation}
where $N_{\rm SL}$ is the number of known and spectroscopically
confirmed strong-lensing clusters, the $dV_\mu/dz$ term denotes the
volume per unit redshift behind an example galaxy cluster that is
magnified by $\mu=\mu_{\rm GW}$, and the last term on the right hand
side is now specific to BBH mergers.

There are 130 spectroscopically confirmed strong-lensing clusters
known at the present time (Section~\ref{sec:lenses}; i.e.\ $N_{\rm
  SL}=130$), with a median redshift of $z=0.3$.  We therefore adopt
the cluster 1E\,0657$-$558 (also known as the ``Bullet Cluster'') at
$z=0.296$ as the example cluster upon which our calculations are
based.  We compute $dV_\mu/dz$ for this cluster using the detailed
parametric mass model of this cluster \citep{Paraficz2016}, and also
confirm that our calculation is numerically converged -- i.e.\ the
volume calculation is insensitive to the width of the redshift bins
that we adopt.  

In principle, the choice of $dn_{\rm BBH}/dV/dt$ involves a circular
argument, given that (1) the current estimate in the local universe is
$12-213\,\perGpcyr$, assuming $\mu=1$ for all BBH GW sources to date
\citep{Abbott2017-GW170104}, and (2) our goal is to explore the
possibility that one or more of the sources were multiply-imaged and
thus at higher redshift.  To break the circularity implied by adopting
the published BBH merger rate for our calculations, we compute the
comoving volume at $0.35<z<2$ that is magnified by $\mu=\mu_{\rm
  GW}(z)$ by the Bullet cluster, and express this as a fraction of the
total comoving volume in this redshift range: $f_\mu\simeq10^{-10}$.
The probability of the GW sources detected to date being
multiply-imaged is therefore very small, and we can safely adopt the
published LIGO BBH merger rate as the local rate, and thus effectively
explore the possibility that one of the detections is multiply-imaged.
Note that $f_\mu\simeq10^{-10}$ is qualitatively consistent with
\citeauthor{Hilbert2008}'s (\citeyear{Hilbert2008}) predictions of the
source frame optical depth to high magnifications.

We now estimate $R_{\rm detect}$, assuming that the BBH merger rate
does not evolve with redshift, and performing the integral in
Equation~\ref{eqn:ratedetect} over the redshift range $0.35<z<2$.  The
results of the calculation are insensitive to this choice.  In
particular we note that $z=0.35$ is $15,000\,{\rm km/s}$ beyond the
cluster redshift and that the cross-section to strong-lensing so close
to the cluster is negligible.  Also, varying the upper limit between
$z=1$ and $z=3$ changes the result negligibly due to the tiny volume
magnified by the very large factors required to push the GW source
redshifts back to $z>1$.  We obtain a rate of detections of BBHs
multiply-imaged by a known and spectroscopically confirmed
strong-lensing cluster per detector year of $R_{\rm
  detect}\simeq7\times10^{-7}-10^{-5}\,{\rm yr}^{-1}$, based on
$dn_{\rm BBH}/dV/dt=12-213\,{\rm Gpc^{-3}yr^{-1}}$.  Adopting a single
and constant value of $dn_{\rm BBH}/dV/dt=50\,\perGpcyr$ yields
$R_{\rm detect}\simeq3\times10^{-6}\,{\rm yr}^{-1}$.  We also consider
a BBH merger rate that tracks the star formation history of the
universe, as described by the fitting function in equation~15 of
\citet{Madau2014}.  This evolving BBH merger rate yields a $R_{\rm
  detect}\simeq4\times10^{-6}-6\times10^{-5}\,{\rm yr}^{-1}$, with a
rate of $R_{\rm detect}\simeq10^{-5}\,{\rm yr}^{-1}$ based on a local
rate of $dn_{\rm BBH}/dV/dt=50\,\perGpcyr$ that evolves as discussed
above.

In summary, based on the calculations detailed above, a reasonable
estimate of the number of BBH mergers detected in LIGO's first two
observing runs and multiply-imaged by a known and spectroscopically
confirmed strong-lensing cluster per detector year is $R_{\rm
  detect}\simeq10^{-5}\,{\rm yr^{-1}}$.  This implies a small and
non-zero probability, that is equivalent to saying that if one of the
BBH detections to date has been multiply-imaged, then this implies
getting lucky at the level of a $\sim4\sigma$ outlier per detector
year.  This is a lower limit on the rate and an upper limit on the
significance at which the the hypothesis of strong-lensing by a
cluster can be ruled out, because our calculations are based on known
strong lensing clusters, and not the full population of clusters that
have sufficiently dense cores to be able to produce strong-lensing
effects.  The low probability of strong-lensing is due to a
combination of (1) the large magnification factors required to
reinterpret the strain signal as coming from a redshift beyond the
population of known lenses, (2) constraints on the local BBH
merger rate are already stringent enough to ensure that $R_{\rm
  detect}\ll1$ at the present time, and (3) the physically plausible
redshift evolution is not strong enough and the luminosity function is
not steep enough to allow lensed sources to dominate the early
detections \citep[see also][]{Dai2017,Ng2017}.  Nevertheless, the rate
is non-zero, and thus the possibility of parameter mis-estimation for
apparently heavy BHs remains.  In the next section we therefore
investigate whether any known cluster strong lenses are consistent
with the GW sky localisations.


\section{The cluster lenses}\label{sec:lenses}

We have assembled a list of 130 spectroscopically confirmed strong
cluster lenses from the literature, drawing mainly on \emph{HST}
studies of X-ray selected clusters, and strong-lensing clusters from
the Sloan Digital Sky Survey \cite[][]{Smith05a, Smith09,
  Limousin07a1689, Limousin2012, Richard10, Christensen2012,
  Oguri2012, Jauzac2015, Umetsu2016, Paraficz2016}.  We compare the
celestial coordinates of these clusters with the sky localisation maps
of the BBH merger detections from 2015, finding that none of the known
strong-lensing clusters are located within the 90\% credible sky
localisation of GW150914, two are located within the 90\% credible
region of GW151226 (MACS\,J0140.0$-$0555 at $z=0.451$, and
MACS\,J1311.0$-$0310 at $z=0.398$), and one is within the 90\%
credible region of LVT151012 (RCS0224$-$0002 at $z=0.773$).  We note
that there are no clusters in common between the sky localisations of
GW151226 and LVT151012, and none of them lie in or near the
intersection of the sky localisations of these two events \citep[see
  for example, Figure~6 of][]{BBH-O1}.  Detailed mass models are
available for all three clusters
\citep{Christensen2012,Ho2012,Smit2017}.

\begin{figure}
  \centerline{
    \includegraphics[width=0.7\hsize,angle=-90]{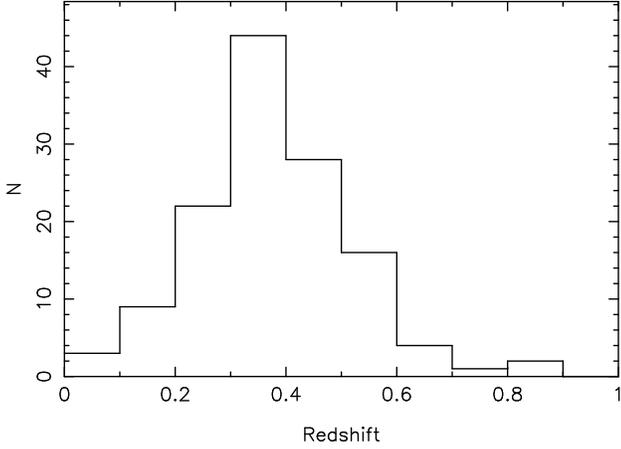}
  }
  \caption{The redshift distribution of the 130 spectroscopically
    confirmed strong-lensing clusters discussed in
    Section~\ref{sec:lenses}.}
  \label{fig:zhist}
\end{figure}

The sky localisations of all three GW events intersect the disk of the
Milky Way.  Unfortunately, severe dust extinction and stellar
obscuration make it very difficult to find clusters at low galactic
latitude, let alone identify whether any of them are strong-lenses.
Despite some clusters being identified at low galactic latitude
\citep[e.g.][]{Ebeling2002, Kocevski2007}, searches for strong-lensing
clusters have concentrated on high latitudes ($|b|>20^\circ$).  It is
therefore possible that an unknown massive galaxy cluster at low
galactic latitude strongly-lensed one or more of the GW events.  This
underlines the fact that the rate of $R_{\rm
  detect}\simeq10^{-5}\,{\rm yr}^{-1}$ (Section~\ref{sec:glgw}), and the
numbers of strong-lensing clusters that we find within the 90\%
credible sky localisations are all lower limits on the incidence of
strong-lensing.


\begin{figure*}
  \centerline{
    \includegraphics[width=0.42\hsize,angle=-90]{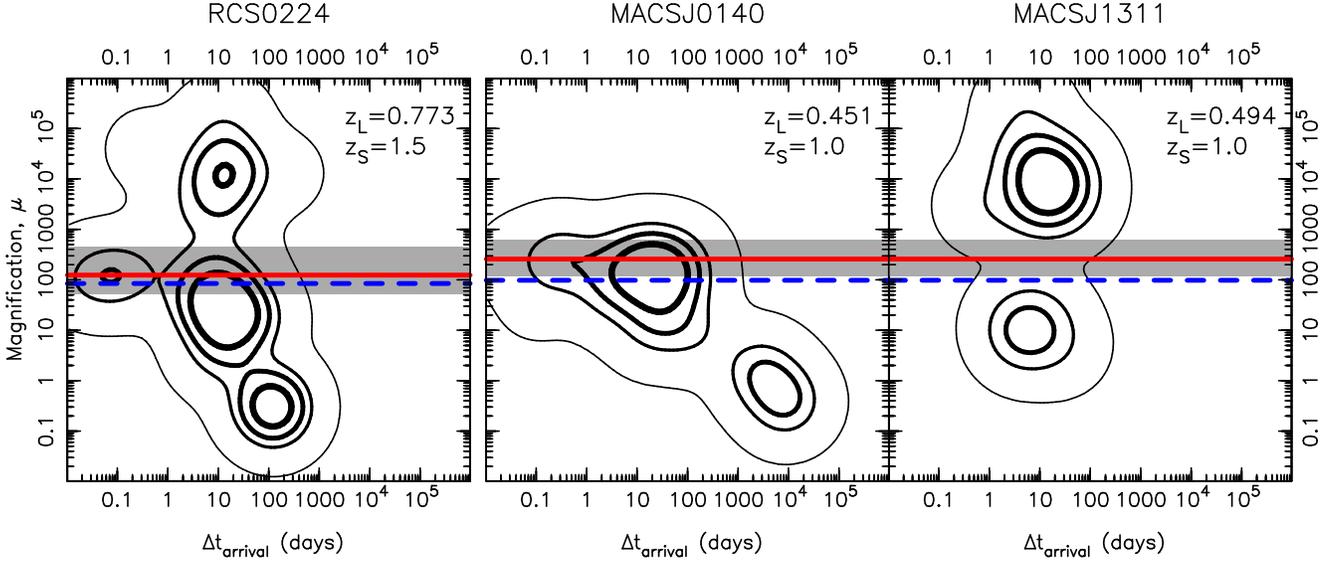}
  }
  \caption{Magnification and time delay suffered by the putative next
    appearance of LVT151012 (left) and GW151226 (centre and right) in
    the scenario each has been multiply-imaged by a strong-lensing
    cluster located in their respective 90\% credible sky
    localisations.  The solid red line shows $\mu_{\rm GW}$, and the
    grey band shows the range $\mu_{\rm GW}^-<\mu<\mu_{\rm GW}^+$.
    The blue dashed lines show $\mu_{\rm GW}/\alpha^2$, the
    magnification threshod above which the next appearance would be
    detectable by LIGO at ${\rm SNR}\gs8$.  Contours enclose 25\%,
    50\%, 75\%, and 99\% of the probability density.  Further details
    are discussed in Section~\ref{sec:calc}.}
  \label{fig:timemag}
\end{figure*}

The detections of GW170104 and GW170814 were announced during the
latter stages of preparing this article, and while responding to the
referee report respectively
\citep{Abbott2017-GW170104,Abbott2017-GW170814}.  We identified two
cluster lenses within the 90\% credible region of both of these
detections.  These findings do not alter any of the conclusions and
discussion presented here.  We will present our follow-up observations
of clusters related to GW170814 in a future article.

\section{Time delay and magnification calculations}\label{sec:calc}

We estimate the arrival times and magnifications of putative future
appearances of GW151226 and LVT151012 due to the three cluster lenses
discussed in Section~\ref{sec:lenses}.  The detailed mass models
referred to above are all constrained by spectroscopically confirmed
multiply-imaged galaxies, thus breaking the redshift space
degeneracies.  The mass distribution of each cluster core was modelled
as a superposition of mass components that represent the large-scale
cluster mass distribution, and the cluster galaxies, and optimized
using the publicly available \textsc{Lenstool} software
\citep{Jullo07}, following methods initially developed by
\citet{Kneib96} and \citet{Smith05a}.  Full details of the models are
presented in \citet{Christensen2012}, \citet{Ho2012}, and
\citet{Smit2017}.

Starting from these models, we identify the sky locations in the
$z_{\rm S}=1$ and $z_{\rm S}=1.5$ source planes of each cluster that
are magnified by $\mu_{\rm GW}^-<\mu<\mu_{\rm GW}^+$, where $\mu_{\rm
  GW}^-$ and $\mu_{\rm GW}^+$ are the values of $\mu_{\rm GW}$ implied
by the lower and upper 90\% confidence intervals respectively on the
unlensed luminosity distance to the sources.  Then we ray traced
these sky locations through the relevant lens models to obtain the
respective image positions, $\vec{\theta}$.  Given the large
magnification values, all of these sky locations are multiply-imaged.
We then measured the gravitational potential at the image positions,
$\phi({\vec{\theta}})$.  The arrival-time surface for a light ray
emitted by a lensed source, at the source-plane position
$\vec{\beta}$, traversing the cluster lens at the image-plane position
$\vec{\theta}$, is given by:
 \begin{equation}
 \tau(\vec{\theta},\vec{\beta}) = \frac{1 + z_{\rm L}}{c} \frac{D_{\rm
    OL} D_{\rm OS}}{D_{\rm LS}} \left[\frac{1}{2} (\vec{\theta} -
  \vec{\beta})^{2} - \phi(\vec{\theta}) \right]
\label{timedel_eq}
\end{equation}
where, $c$ is the speed of light in vacuum, $z_{\rm L}$ is the
redshift of the cluster lens, $D_{\rm OL}$, $D_{\rm OS}$, and $D_{\rm
  LS}$ are the observer-lens, observer-source, and lens-source angular
diameter distances respectively, and $\phi(\vec{\theta})$ represents
the projected cluster gravitational potential \citep{Schneider1985}.
These calculations were performed following the analytic procedure
described by \cite{Jauzac2016}.

The distribution of time delay between the first arrival of an image
that suffers $\mu_{\rm GW}^-<\mu<\mu_{\rm GW}^+$ and the next arrival
of an image from the same source location that is detectable by LIGO,
$\Delta{t}_{\rm arrival}$, spans a fraction of a day to a few years
(Figure~\ref{fig:timemag}).  We classify these ``next images'' as
being detectable if they are magnified by $\mu\ge\mu_{\rm
  GW}/\alpha^2$ where $\alpha=13/8$, and $9.7/8$ for GW151226, and
LVT151012 respectively, i.e.\ the ratio of the SNR at which each was
detected in 2015 and the minimum SNR required of a detection by LIGO
\citep{LIGOdetect}.  Based on these calculations, we estimate the
fraction of the next images that would be detectable by LIGO is
$\sim20-60\%$.  Note that in Figure~\ref{fig:timemag} we show results
for $z_{\rm S}=1$ for the clusters at $z_{\rm L}<0.5$, in order to
restrict our attention to values of $\mu_{\rm GW}$ that not too
extreme.  However, we show results for $z_{\rm S}=1.5$ for
RCS0224$-$0002, because the cross-section of this high-redshift
cluster to a source at $z_{\rm S}=1$ is tiny.  Our results are
insensitive to these choices of $z_{\rm S}$.


\section{Summary and discussion}\label{sec:summary}

The degeneracy between gravitational magnification ($\mu$) and
luminosity distance \citep{Wang1996} causes the luminosity distance to
a GW source to be revised upwards by a factor of $\sqrt{\mu}$ if it is
gravitationally magnified, and the inferred source frame masses of the
compact objects to be revised down by a factor of $(1+z)$.  This is
interesting because some of the early GW detections appear to come
from heavy BHs \citep{Interpretation150914,Stevenson2017}, and
gravitational lensing by massive galaxy clusters have been central to
the first detection of distant galaxy populations, most notably at
sub-mm wavelengths \citep{Smail97,Ivison98}.

We estimate, based on the known populations of BBH GW sources and
strong-lensing clusters, that the number of BBH mergers detected per
detector year at LIGO/Virgo's current sensitivity that are
multiply-imaged by known and spectroscopically confirmed
strong-lensing clusters is $R_{\rm detect}\simeq10^{-5}\,{\rm
  yr}^{-1}$.  This calculation takes into account the gravitational
optics of clusters with sky positions consistent with the GW
detections, the local BBH merger rate, and the star formation history
of the universe.  It is a conservative lower limit on the true rate of
multiply-imaged GWs to date, because it is based only on the known
lenses, and in particular ignores any strong lenses obscured by the
Galactic Plane.  The rate that we have computed is equivalent to
saying that the hypothesis that one of the BBH GWs detected to date
was multiply-imaged can be rejected at $\ls4\sigma$.  It is therefore
unlikely, but not impossible that one of the GWs detected to date was
multiply-imaged.

Our search for candidate strong-lensing clusters that might have
multiply-imaged GW sources concentrates on the BBH detections in
LIGO's first observing run.  Based on a comparison of the celestial
coordinates of 130 spectroscopically confirmed strong-lensing clusters
with the GW sky localisation maps, we have identified no candidate
lenses within the 90\% credible sky localisation of GW150914, two
within the 90\% credible sky localisation of GW151226
(MACS\,J0410.0$-$0555 and MACS\,J1311.0$-$0311), and one within the
90\% credible sky localisation of LVT151012 (RCS0224$-$0002).  We used
detailed mass models of these three clusters to calculate the
magnifications and time delays suffered by the putative next
appearance of GW151226 and LVT151012, in the scenario that they have
indeed been multiply-imaged.  We find that $20-60\%$ of the next
appearances would be detectable by LIGO/Virgo at the sensitivity
achieved in the first and second observing runs, and that they would
arrive at Earth within $3\,{\rm years}$ of the original detections.

Finally, we consider what it would take to identify unambiguously a
multiply-imaged GW.  Identifying a temporally coincident optical
transient in the strong-lensing region of a massive galaxy cluster
located within the GW sky localisation would be an ideal scenario.
This would allow the previous and subsequent appearances of the
optical transient and presumed associated GW source to be computed,
based on a detailed model of the cluster mass distribution.
Predictions of previous appearances could then be compared with
earlier GW detections and archival optical observations, and
predictions of future appearances would inform future observations.
Based on this outline, the gold standard would therefore be temporal
and celestial sphere coincidence of the sky localisations of two GW
detections and two optical transients with the strong-lensing region
of a single cluster lens.  Moreover, consistency between the strain
signals detected by LIGO/Virgo would be required; this is efficently
phrased as requiring consistency between the detector frame chirp
masses of the two GW detections.  Chirp masses are currently measured
to few per cent precision \citep[e.g.][]{Abbott2017-GW170814}.

Given the sensitivity and thus reach of LIGO/Virgo, it is more
realistic to contemplate searching for multiply-imaged GW sources that
include one, preferrably two BHs.  Whilst it is an open question as to
whether BBH mergers emit any EM radiation, it is reasonable to expect
that any optical emission will be faint, notwithstanding any boost to
the flux level thanks to gravitational magnification.  It is therefore
appropriate to consider deep follow-up optical observations of
candidate strong-lensing clusters located in the sky localisations of
BBH and BHNS GW sources, and also to consider whether and how a
multiply-imaged GW source might be identified without an EM
counterpart.  On the first point, we commenced deep follow-up
observations of candidate strong-lensing clusters in the latter stages
of LIGO's second observing run with MUSE on VLT and GMOS on
Gemini-South.  These observations aim to reach a depth of ${\rm
  AB}\simeq25$ per epoch, which is considerable deeper than typical
observations of GW sky localisations with wide-field instruments.  We
will report on these follow-up observations in a future article.  On
the latter point, we intend to explore the feasibility of basing the
discovery of multiply-imaged GW sources on solely the sky
localisations, chirp masses, and available strong-lensing clusters for
a given pair of GW detections.

\section*{Acknowledgments}

We thank an anonymous referee for comments that helped us to improve
this article.  GPS thanks Alberto Vecchio for numerous stimulating
discussions on strong-lensing of gravitational waves.  We also thank
Marceau Limousin, Christopher Berry, Keren Sharon, Danuta Paraficz,
and Ilya Mandel for assistance and comments.  GPS, MJ, and JR thank
Jean-Paul Kneib and Eric Jullo for creating {\sc Lenstool} and
teaching us how to use it.  We acknowledge support from the Science
and Technology Facilities Council through the following grants:
ST/N000633/1 (GPS, WMF), ST/P000541/1 (MJ), and ST/K005014/1 (JV).  MJ
also acknowledges support from the ERC advanced grant LIDA, and RM
acknowledges support from the Royal Society.  This work used the LIGO
Open Science Centre data releases for GW150914, GW151226, and
LVT151012, with DOI 10.7935/K5MW2F23, 10.7935/K5H41PBP, and
10.7935/K5CC0XMZ respectively.

\bibliographystyle{mn2e}

\label{lastpage}

\end{document}